\documentclass[pra,showpacs]{revtex4}
\usepackage{graphicx}
\usepackage{amsmath}
\usepackage{amssymb}
\input{epsf}

\newcommand{\rem}[1]{}

\begin{document}

\title{Quantum computing and information extraction for a dynamical 
quantum system}
\author{Giuliano Benenti}
\email{giuliano.benenti@uninsubria.it}
\homepage{http://www.unico.it/~dysco}
\affiliation{Center for Nonlinear and Complex Systems, Universit\`a degli 
Studi dell'Insubria, Via Valleggio 11, 22100 Como, Italy}
\affiliation{Istituto Nazionale per la Fisica della Materia, 
Unit\`a di Como, Via Valleggio 11, 22100 Como, Italy}
\author{Giulio Casati}
\email{giulio.casati@uninsubria.it}
\affiliation{Center for Nonlinear and Complex Systems, Universit\`a degli 
Studi dell'Insubria, Via Valleggio 11, 22100 Como, Italy}
\affiliation{Istituto Nazionale per la Fisica della Materia, 
Unit\`a di Como, Via Valleggio 11, 22100 Como, Italy}
\affiliation{Istituto Nazionale di Fisica Nucleare,
Sezione di Milano, Via Celoria 16, 20133 Milano, Italy}
\author{Simone Montangero}
\email{monta@sns.it}
\homepage{http://www.sns.it/~montangero}
{\affiliation{NEST-INFM $\&$ Scuola Normale Superiore, 
Piazza dei Cavalieri 7, 56126 Pisa, Italy}
\date{January 30, 2004}
\pacs{03.67.Lx, 05.45.Mt}

\begin{abstract} 
We discuss the simulation of a complex dynamical system, 
the so-called quantum sawtooth map model, on a quantum computer. 
We show that a quantum computer can be used to efficiently extract 
relevant physical information for this model.
It is possible to simulate the dynamical localization of classical
chaos and extract the localization length of the system with 
quadratic speed up with respect to any known classical computation.
We can also compute with algebraic speed up 
the diffusion coefficient and the diffusion exponent 
both in the regimes of Brownian and anomalous diffusion.
Finally, we show that it is possible to extract the fidelity of the quantum 
motion, which measures the stability of the system under perturbations,
with exponential speed up. 
\end{abstract}
\maketitle

\section{Introduction}
\label{sec:introduction}

One of the main applications of computers is the simulation of dynamical 
models describing the evolution of complex systems. From the viewpoint
of quantum computation, quantum mechanical systems play a special 
role. Indeed, the simulation of quantum many-body problems on a 
classical computer is a difficult task as the size of the Hilbert space 
grows exponentially with the number of particles. For instance, if we 
wish to simulate a chain of $n$ spin-$\frac{1}{2}$ particles, the size of 
the Hilbert space is $2^n$. Namely, the state of this
system is determined by $2^n$ complex numbers. As observed by Feynman
in the 1980's \cite{feynman}, the growth in memory requirement is only
linear on a quantum computer, which is itself a many-body quantum system. For
example, to simulate $n$ spin-$\frac12$ particles we only need $n$
qubits. Therefore, a quantum computer operating with only a few tens of
qubits could outperform a classical computer. 
More recently, a few quantum efficient algorithms have been developed
for various quantum systems, ranging from some many-body problems 
\cite{lloyd,fermions} to single-particle models of quantum chaos 
\cite{schack,georgeot,bcms01}.

Any quantum algorithm has to address the problem of efficiently 
extracting useful information from the quantum computer wave function.
The result of the simulation of a quantum system is the wave 
function of this system, encoded in the $n$ qubits of the quantum 
computer. The problem is that, in order to measure all $N=2^n$ 
wave function coefficients by means of standard polarization 
measurements of the $n$ qubits, one has to repeat the quantum
simulation a number of times exponential in the number of qubits. 
This procedure would spoil any quantum algorithm, even in the 
case in which such algorithm could compute the wave function 
with an exponential gain with respect to any classical computation.
Nevertheless, there are some important physical questions that can be 
answered in an efficient way, and we will discuss a few examples 
in this paper.

We will discuss a quantum algorithm which efficiently simulates 
the quantum sawtooth map, a physical model with rich and complex dynamics
\cite{bcms01}. This system is characterized by very different
dynamical regimes, ranging from integrability to chaos, and 
from normal to anomalous diffusion; it also
exhibits the phenomenon of dynamical localization of classical 
chaotic diffusion. We will show that some important physical 
quantities can be extracted efficiently by means of a quantum computer: 

(i) the \emph{localization length} of the system, which can be 
extracted with a \emph{quadratic speed up} with respect 
to any known classical computation \cite{bcms03};

(ii) the \emph{diffusion coefficient} and the \emph{diffusion 
exponent}, both in the regimes of normal (Brownian) and 
anomalous diffusion; in this case we obtain an
\emph{algebraic speed up};

(iii) the \emph{fidelity} of quantum motion, 
which characterizes the stability 
of the system under perturbations; for this quantity we 
achieve an \emph{exponential speed up}.  

The paper is organized as follows: 
the properties of the sawtooth map model are discussed in
Sec.~\ref{sec:sawmap}; 
our quantum algorithm simulating the quantum dynamics of this model
in Sec.~\ref{sec:algorithm};
the quantum computation of the localized regime and the extraction 
of the localization length in Sec.~\ref{sec:localization};
the quantum simulation of the phenomena of normal and anomalous
diffusion and the computation of the diffusion coefficient 
and diffusion exponent in Sec.~\ref{sec:anomalous};
the quantum computation of the fidelity of quantum motion in 
Sec.~\ref{sec:fidelity};
our conclusions are summarized in Sec.~\ref{sec:conclusions}.

\section{The sawtooth map}
\label{sec:sawmap}

The sawtooth map is a prototype model in the studies of classical
and quantum-dynamical systems and exhibits a rich variety of
interesting physical phenomena, from complete chaos to complete integrability,
normal and anomalous diffusion, dynamical localization, and cantori
localization. Furthermore, the sawtooth map gives a good approximation
to the motion of a particle bouncing inside a stadium billiard (which
is a well-known model of classical and quantum chaos). 

The sawtooth map belongs to the class of 
periodically driven dynamical
systems, governed by the Hamiltonian
\begin{equation}
  H(\theta,I;\tau) =
  \frac{I^2}{2} + V(\theta) 
  \sum_{j=-\infty}^{+\infty} \delta(\tau-jT) \,,
  \label{sawham}
\end{equation}
where $(I,\theta)$ are conjugate action-angle variables
($0\leq\theta<2\pi$). This Hamiltonian is the sum of two terms,
$H(\theta,I;\tau)=H_0(I)+U(\theta;t)$, where $H_0(I)=I^2\!/2$
is just the kinetic energy of a free rotator (a particle moving on 
a circle parametrized by the coordinate $\theta$), while
\begin{equation}
  U(\theta;t) =
  V(\theta) \sum_j \delta(\tau-jT)
\end{equation}
represents a force acting on the particle 
that is switched on and off instantaneously at time intervals
$T$. Therefore, we say that the dynamics described by Hamiltonian
(\ref{sawham}) is \emph{kicked}. 
The corresponding Hamiltonian
equations of motion are 
\begin{equation}
  \left\{
    \begin{array}{l}
      \displaystyle
      \dot{I} =
      -\frac{\partial{H}}{\partial\theta} =
    -\frac{d V(\theta)}{d \theta}
    \sum_{j=-\infty}^{+\infty} \delta(\tau-jT) \,,
    \\[2ex]
      \displaystyle
      \dot\theta =
      \frac{\partial{H}}{\partial{I}} = I \,.
    \end{array}
  \right.
\end{equation}
These equations can be easily integrated and one finds that the evolution from
time $lT^-$ (prior to the $l$-th kick) to time $(l+1)T^-$
(prior to the $(l+1)$-th kick) is described by the map
\begin{equation}
  \left\{
    \begin{array}{l}
      \displaystyle
      \bar{I} = I + F (\theta) \,,
    \\[2ex]
      \displaystyle
      \bar\theta= \theta + T\bar{I} \,,
    \end{array}
  \right.
  \label{sawmap}
\end{equation}
where $F(\theta)=-dV(\theta)/d\theta$ is the force acting on 
the particle.

In the following, we will consider the special case 
$V(\theta)=-k(\theta-\pi)^2/2$. 
This map is called the \emph{sawtooth map}, since the force
$F(\theta)=-dV(\theta)/d\theta=k(\theta-\pi)$ has a sawtooth shape, 
with a discontinuity at $\theta=0$. By rescaling $I\to{J=TI}$, 
the classical dynamics is seen to depend only on the
parameter $K=kT$. Indeed, in terms of the variables $(J,\theta)$ map
(\ref{sawmap}) becomes
\begin{equation}
  \left\{
    \begin{array}{l}
      \displaystyle
      \bar{J} = J + K (\theta - \pi) \,,
    \\[2ex]
      \displaystyle
      \bar{\theta} = \theta + \bar{J} \,.
    \end{array}
  \right.
  \label{sawmap2}
\end{equation}
The sawtooth map exhibits sensitive dependence 
on initial conditions, which is the
distinctive feature of classical chaos: any small error 
is amplified exponentially in time. In other
words, two nearby trajectories separate exponentially, with a rate given 
by the maximum Lyapunov exponent $\lambda$, defined as 
\begin{equation}
  \lambda = 
  \lim_{|t|\to\infty} \frac1{t}
  \ln\!\left( \frac{\delta(t)}{\delta(0)} \right) ,
\end{equation}
where the discrete time $t=\tau/T$ measures the number of map 
iterations and
$\delta(t)=\sqrt{[\delta J(t)]^2+[\delta \theta(t)]^2}$. To
compute $\delta J(t)$ and $\delta \theta(t)$, we differentiate map 
(\ref{sawmap2}), obtaining
\begin{equation}
  \left[
    \begin{array}{c}
      \delta \bar{J} \\
      \delta \bar\theta
    \end{array}
  \right]
  = M
  \left[
    \begin{array}{c}
      \delta J \\
      \delta\theta
    \end{array}
  \right]
  =
  \left[
    \begin{array}{c@{\quad}c}
      1 &  K  \\
      1 & 1+K
    \end{array}
  \right]
  \left[
    \begin{array}{c}
      \delta J \\
      \delta\theta
    \end{array}
  \right] .
\label{tangmap}
\end{equation}
The iteration of map (\ref{tangmap}) gives $\delta J(t)$ and
$\delta \theta(t)$ as a function of $\delta J(0)$ and 
$\delta \theta(0)$ [$\delta J(0)$ and $\delta \theta(0)$ 
represent a change of the initial conditions].
The stability matrix $M$ has eigenvalues
$\mu_{\pm}=\frac{1}{2}(2+K\pm\sqrt{K^2+4K})$, which do not depend
on the coordinates $J$ and $\theta$ and are complex conjugate
for $-4\leq{K}\leq0$ and real for $K<-4$ and $K>0$. 
Thus, the classical motion is stable for $-4\leq{K}\leq0$ 
and completely chaotic for $K<-4$ and $K>0$.
For $K>0$, $\delta (t) \propto (\mu_+)^t$ asymptotycally 
in $t$, and therefore the maximum Lyapunov exponent 
is $\lambda=\ln\mu_+$. Similarly, we obtain 
$\lambda=\ln|\mu_-|$ for $K<-4$. In the stable region
$-4\le{K}\le0$, $\lambda=0$.

The sawtooth map can be studied on the cylinder [$J\in(-\infty,+\infty)$],
or on a torus of sinite size ($-{\pi}L\le J < \pi L$, where $L$ is an
integer, to assure that no discontinuities are introduced in the second
equation of (\ref{sawmap2}) when $J$ is taken modulus $2{\pi}L$). 
Although the sawtooth map is a deterministic system, for $K>0$ and $K<-4$
the motion of a trajectory along the momentum direction is in practice 
indistinguishable from a random walk. Thus, one has normal diffusion in 
the action (momentum) variable and the evolution of the distribution 
function $f(J,t)$ is governed by a Fokker--Planck equation:
\begin{equation}
  \frac{\partial{f}}{\partial{t}} =
  \frac{\partial}{\partial{J}}
  \left(\frac12 D \frac{\partial{f}}{\partial{J}} \right) .
  \label{fokkerplanck}
\end{equation}
The diffusion coefficient $D$ is defined by
\begin{equation}
  D = \lim_{t\to\infty} \frac{\langle(\Delta{J}(t))^2\rangle}{t} \,,
\end{equation}
where $\Delta{J}\equiv{J}-\langle{J}\rangle$, and $\langle\dots\rangle$ denotes
the average over an ensemble of trajectories. If at time $t=0$ we take 
a phase space distribution with initial momentum $J_0$ and random phases
$0\leq\theta<2\pi$, then the solution of the Fokker--Planck equation
(\ref{fokkerplanck}) is given by
\begin{equation}
  f(J,t) =
  \frac1{\sqrt{2 \pi D t}} \, \exp\!\left[ -\frac{(J-J_0)^2}{2Dt} \right] .
\end{equation}
The width $\sqrt{\langle(\Delta J(t))^2\rangle}$
of this Gaussian distribution grows in time, according to
\begin{equation}
 \langle (\Delta{J}(t))^2 \rangle \approx 
 D(K) \, t \,.
\end{equation}
For $K>1$, the diffusion coefficient is well approximated by the random phase
approximation, in which we assume that there are no correlations between the
angles (phases) $\theta$ at different times. Hence, we have
\begin{equation}
  D(K) \approx
  \langle   (\Delta{J}_1)^2 \rangle = 
  \frac1{2\pi}\int_0^{2\pi} d\theta \, (\Delta{J}_1)^2  =
  \frac1{2\pi}\int_0^{2\pi} d\theta \, K^2(\theta-\pi)^2 =
  \frac{\pi^2}{3} \, K^2 \,,
\end{equation}
where $\Delta{J}_1=\bar{J}-J$ is the change in action after a single map step.
For $0<K<1$ diffusion is slowed, due to the sticking of trajectories
close to broken tori (known as cantori), and we have
$D(K)\approx3.3\,K^{5/2}$ (this regime is discussed in \cite{percival}). For
$-4<K<0$ the motion is stable, the phase space has a complex structure of
elliptic islands down to smaller and smaller scales, and one can observe
anomalous diffusion, that is,
$\langle(\Delta{J})^2\rangle\propto{t}^\alpha$, with $\alpha\ne1$ (for instance,
$\alpha=0.57$ when $K=-0.1$, see Fig.~\ref{fanom} below). 
The cases $K=-1,-2,-3$ are integrable.

The quantum version of the sawtooth map
is obtained by means of the usual
quantization rules, $\theta\to\hat{\theta}$ and 
$I\to{}\hat{I}=-i\partial/\partial\theta$
(we set $\hbar=1$). The quantum evolution in one map iteration is described by
a unitary operator $\hat{U}$, called the Floquet operator, 
acting on the wave function $\psi$:
\begin{equation}
  \bar\psi =
  \hat{U}\,\psi =
  \exp\left[
    -i \int_{lT^-}^{(l+1)T^-} d\tau H(\hat{\theta},\hat{I};\tau)
  \right]
  \psi \,,
  \label{sawq}
\end{equation}
where $H$ is Hamiltonian (\ref{sawham}). Since the potential $V(\theta)$ is
switched on only at discrete times $lT$, it is straightforward to
obtain
\begin{equation}
  \bar\psi =
  e^{-i T \hat{I}^2\!/2} \, e^{-iV(\hat{\theta})} \, \psi =
  e^{-i T \hat{I}^2\!/2} \, e^{ik(\hat{\theta} -\pi\hat{\openone})^2\!/2} 
\, \psi \,,
  \label{sawquantum}
\end{equation}
where $\hat{\openone}$ denotes the identity operator.
It is important to emphasize that, while the classical 
sawtooth map depends only on the rescaled parameter $K=kT$, the 
corresponding quantum evolution (\ref{sawquantum}) depends on 
$k$ and $T$ separately. 
The effective Planck constant is given by $\hbar_\text{eff}=T$.
Indeed, if we consider the operator $\hat{J}=T\hat{I}$ 
($\hat{J}$ is the quantization of the classical rescaled action $J$),
we have 
\begin{equation}
[\hat{\theta},\hat{J}]=T[\hat{\theta},\hat{I}]=i T =i \hbar_\textbf{eff}.
\end{equation} 
The classical limit $\hbar_\text{eff}\to 0$ is obtained by taking 
$k\to\infty$ and $T\to0$, while keeping $K=kT$ constant.

\section{Quantum computing of the quantum sawtooth map}
\label{sec:algorithm}

In the following, we describe an exponentially efficient quantum algorithm for
simulation of the map (\ref{sawquantum}) \cite{bcms01}.
It is based on the forward/backward quantum Fourier transform between momentum
and angle bases. Such an approach is convenient since the operator $\hat{U}$,
introduced in Eq.~(\ref{sawq}), is the product of two operators,
$\hat{U}_k=e^{ik(\hat{\theta}-\pi\hat{\openone})^2\!/2}$ and 
$\hat{U}_T=e^{-iT\hat{I}^2\!/2}$,
diagonal in the $\theta$ and $I$ representations, respectively. This quantum
algorithm requires the following steps for one map iteration:
\begin{enumerate}
\item
We apply $\hat{U}_k$ to the wave function $\psi(\theta)$. In order to decompose
the operator $\hat{U}_k$ into one- and two-qubit gates, we first of all
write $\theta$ in binary notation:
\begin{equation}
  \theta=2\pi\sum_{j=1}^n \alpha_j 2^{-j} \,,
\end{equation}
with $\alpha_i\in \{ 0,1 \}$. Here $n$ is the number of qubits, so that the
total number of levels in the quantum sawtooth map is $N=2^n$. From this
expansion, we obtain
\begin{equation}
  (\theta - \pi)^2 =
  4\pi^2 \sum_{j_1,j_2=1}^n
  \left( \frac{\alpha_{j_1}}{2^{j_1}} -\frac1{2n} \right)
  \left( \frac{\alpha_{j_2}}{2^{j_2}} -\frac1{2n} \right) ,
\end{equation}
that is 
\begin{equation}
  (\hat{\theta} - \pi\hat{\openone})^2 =
  4\pi^2 \sum_{j_1,j_2=1}^n
  \hat{\openone}_1\otimes \cdots \otimes 
  \hat{\openone}_{j_1-1} \otimes
  \hat{O}_{j_1}\otimes
  \hat{\openone}_{j_1+1}\otimes \cdots \otimes 
  \hat{\openone}_{j_2-1} \otimes
  \hat{O}_{j_2}\otimes
  \hat{\openone}_{j_2+1}\otimes \cdots 
  \otimes \hat{\openone}_{j_n},
\label{th2operator}
\end{equation}
where $\hat{\openone}_j$ is the identity operator for the qubit $j$ and 
the one-qubit operators $\hat{O}_{j_1}$ and $\hat{O}_{j_2}$ 
act on qubits $j_1$ and $j_2$, respectively. We have 
\begin{equation}
\hat{O}_{j}=
\frac{1}{2^j}\frac{\hat{\openone}_j-(\hat{\sigma}_z)_j}{2}
-\frac{1}{2n} \hat{\openone}_j,
\end{equation}
where $(\hat{\sigma}_z)_j$ denotes the Pauli operator 
$\hat{\sigma}_z$ for the qubit $j$.
Note that the operator $\hat{O}_j$ is diagonal in the 
computational basis $\{|0\rangle,|1\rangle\}$.
We can insert (\ref{th2operator}) into the unitary operator $\hat{U}_k$, 
obtaining the decomposition
\begin{equation}
  e^{ik(\hat{\theta} -\pi\hat{\openone})^2\!/2} =
  \prod_{j_1,j_2=1}^n
  \exp\!\left[
    i 2 \pi^2 k
    \left(
  \hat{\openone}_1\otimes \cdots \otimes 
  \hat{\openone}_{j_1-1} \otimes
  \hat{O}_{j_1}\otimes 
  \hat{\openone}_{j_1+1}\otimes \cdots \otimes 
  \hat{\openone}_{j_2-1} \otimes
  \hat{O}_{j_2}\otimes
  \hat{\openone}_{j_2+1}\otimes \cdots \otimes 
  \hat{\openone}_{j_n}
  \right)
  \right] ,
  \label{ukdec}
\end{equation}
which is the product of $n^2$ two-qubit gates (controlled phase-shift 
gates), each acting non-trivially only on the
qubits $j_1$ and $j_2$. In the computational basis
$\{|\alpha_{j_1}\alpha_{j_2}\rangle =
|00\rangle,|01\rangle,|10\rangle,|11\rangle\}$ each two-qubit gate can be
written as $\exp(i2\pi^2kD_{j_1,j_2})$, where $D_{j_1,j_2}$ is a diagonal
matrix:
\begin{equation}
  D_{j_1,j_2} =
  \left[
    \begin{array}{cccc}
      \frac1{4n^2} & 0 & 0 & 0 \\
      0 & -\frac1{2n}\big(\frac1{2^{j_2}}-\frac1{2n}\big) & 0 & 0 \\
      0 & 0 & -\frac1{2n}\big(\frac1{2^{j_1}}-\frac1{2n}\big) & 0 \\
      0 & 0 & 0 & \big(\frac1{2^{j_1}}-\frac1{2n}\big)
      \big(\frac1{2^{j_2}}-\frac1{2n}\big)
    \end{array}
  \right] .
\end{equation}
Note that decomposition (\ref{ukdec}) of
$\hat{U}_k$ is specific to the sawtooth map.
\item
The change from the $\theta$ to the $I$ representation is obtained by means of
the quantum Fourier transform, which requires $n$ Hadamard gates
and $\frac12 n(n-1)$ controlled phase-shift gates
(see, e.g., \cite{nielsenchuang}).
\item
In the $I$ representation, the operator $\hat{U}_T$ has essentially the same
form as the operator $\hat{U}_k$ in the $\theta$ representation, and therefore
it can be decomposed into $n^2$ controlled phase-shift
gates, similarly to Eq.~(\ref{ukdec}). 
\item
We return to the initial $\theta$ representation by application
of the inverse quantum Fourier transform.
\end{enumerate}
Thus, overall, this quantum algorithm requires $3n^2+n$
gates per map iteration ($3n^2-n$ controlled phase-shifts and
$2n$ Hadamard gates). This number is to be compared with the
$O(n2^n)$ operations required by a classical computer to simulate
one map iteration by means of a fast Fourier transform. Thus,
the quantum simulation of the quantum sawtooth map dynamics is exponentially
faster than any known classical algorithm. 
Note that the resources required to the quantum computer to simulate 
the evolution of the sawtooth map are only logarithmic in the 
system size $N$. 
Of course, there remains
the problem of extracting useful information from the quantum-computer
wave function. This will be discussed in the subsequent sections.

\section{Quantum computing of dynamical localization}
\label{sec:localization}

Dynamical localization is one of the most interesting phenomena that
characterize the quantum behavior of classically chaotic systems:
quantum interference effects suppress chaotic diffusion in momentum,
leading to exponentially localized wave functions. This phenomenon was first
found and studied  in the quantum kicked-rotator model \cite{izrailev} and has
profound analogies with Anderson localization of electronic transport in
disordered materials \cite{fishman}. 
Dynamical localization has been observed experimentally
in the microwave ionization of Rydberg atoms \cite{koch} 
and in experiments with cold atoms \cite{raizen}.

Dynamical localization can be studied in the sawtooth map model. In this case,
map (\ref{sawquantum}) is studied on the cylinder [$I\in(-\infty,+\infty)$],
which is cut-off at a finite number $N$ of levels due to the finite
quantum (or classical) computer memory. Similarly to other models of quantum
chaos, quantum interference in the sawtooth map leads to suppression
of classical chaotic diffusion after a \emph{break time} $t^\star$. 
For $t>t^\star$, while the classical distribution goes on diffusing, 
the quantum distribution reaches a steady state which
\emph{decays exponentially} over the momentum eigenbasis:
\begin{equation}
  W_m \equiv \big| \langle m | \psi \rangle \big|^2 \approx
  \frac1{\ell} \, \exp\!\left[ -\frac{2|m-m_0|}{\ell} \right] ,
  \label{expdecay}
\end{equation}
with $m_0$ the initial value of the momentum 
(the index $m$ singles out the
eigenstates of $\hat{I}$, that is, $\hat{I}|m\rangle=m|m\rangle$) 
\cite{footnote}. 
Therefore, for
$t>t^\star$ only $\sqrt{\langle(\Delta{m})^2\rangle}\sim\ell$ levels are
populated.

An estimate of $t^\star$ and $\ell$ can be obtained by means of 
the following argument \cite{siberia}.
The localized wave packet has significant 
projection over about $\ell$ basis states,
both in the basis of the momentum eigenstates and in the 
basis of the eigenstates of the Floquet operator $\hat{U}$ 
defined by Eq.~(\ref{sawq}). This operator is unitary and therefore
its eigenvalues can be written as $\exp(i\lambda_i)$, and the 
so-called \emph{quasienenergies} $\lambda_i$ are 
in the interval $[0,2\pi[$. Thus, the mean level spacing 
between ``significant'' quasienergy eigenstates is 
$\Delta E \approx 2\pi / \ell$. The Heisenberg principle tells us
that the minimum time required to the dynamics to resolve 
this energy spacing is given by   
\begin{equation}
t^\star \approx 1/\Delta E \approx \ell.
\label{siberia1}
\end{equation}
This is the break time after which the quantum feature 
of the dynamics reveals. Diffusion up to time $t^\star$
involves a number of levels given by 
\begin{equation}
\sqrt{\langle(\Delta{m})^2\rangle}\approx\sqrt{D_mt^\star}\approx \ell, 
\label{siberia2}
\end{equation}
where $D_m=D/T^2$ is the classical diffusion coefficient, 
measured in number of levels.
The relations (\ref{siberia1}) and (\ref{siberia2}) imply 
\begin{equation}
  t^\star \approx \ell \approx D_m.
  \label{loc1}
\end{equation}
Therefore, the quantum localization length $\ell$ for the average probability
distribution is approximately equal to the classical diffusion coefficient.
For the sawtooth map,
\begin{equation}
  \ell \approx D_m \approx  
 (\pi^2\!/3) k^2 \,.
  \label{loc2}
\end{equation}
Note that the quantum localization can take place on a finite system
only if $\ell$ is smaller than the system size $N$.

In Fig.~\ref{locfig} (taken from \cite{bcms03}), we show that
exponential localization can already be clearly seen 
with $n=6$ qubits.
It is important to stress that in a quantum computer the memory capabilities
grow exponentially with the number of qubits (the number of levels $N$ is equal
to $2^n$). Therefore, already with less than $40$ qubits, one could make
simulations inaccessible to today's supercomputers. 
Fig.~\ref{locfig} shows that the
exponentially localized distribution, appearing at
$t\approx{}t^\star$, is \emph{frozen} in time, apart from quantum fluctuations,
which we partially smooth out by averaging over a few map steps. The freezing
of the localized distribution can be seen from comparison of the
probability distributions taken immediately after $t^\star$ (the full
curve in Fig.~\ref{locfig}) and at a much larger time $t=300\approx25t^\star$
(the dashed curve in the same figure). Here the localization length is
$\ell\approx12$, and classical diffusion is suppressed after a break time
$t^\star\approx\ell\approx{}D_m$, in agreement with estimates
(\ref{loc1})--(\ref{loc2})
[the classical diffusion coefficient is
$D_m\approx(\pi^2\!/3)k^2\approx9.9$].
This quantum computation up to times of
the order of $\ell$ requires a number $N_g\approx 3n^2\ell\sim10^3$ 
of one- or two-qubit quantum gates.

\begin{figure}
  \includegraphics[width=9.0cm,angle=0]{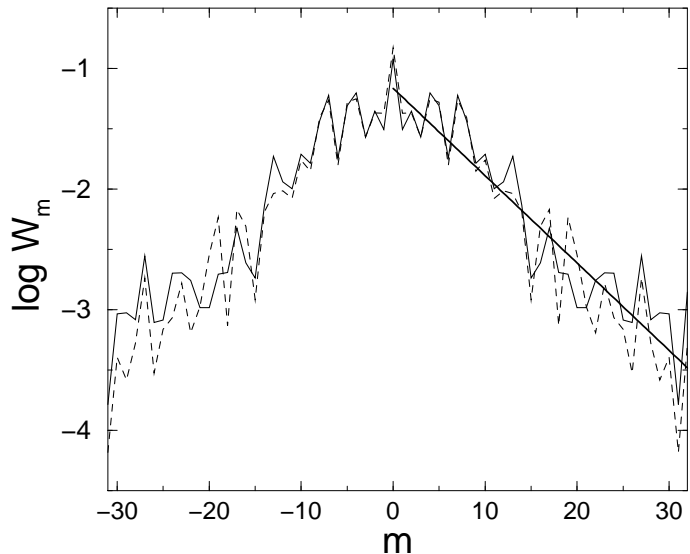}
  \caption{The probability distribution over the momentum basis for the
    sawtooth map with $n=6$ qubits, $k=\sqrt3$, $K=\sqrt2$, and initial
    momentum $m_0=0$; the time average is taken in the intervals
    $10\leq{t}\leq20$ (full curve) and $290\leq{t}\leq300$ (dashed curve). The
    straight line fit, $W_m\propto\exp(-2|m|/\ell)$, gives a localization
    length $\ell\approx12$. Note that the logarithm is base ten.}
  \label{locfig}
\end{figure}

In Fig.~\ref{locfig2}, we show a quantum computation 
that might be performed already with a three-qubit quantum processor. 
It is possible to compare two very different regimes, namely the localized 
and the ergodic regime, by varying only the value of the quantum parameter 
$k$, while keeping the classical parameter $K$ constant. In both cases the 
wave function is stationary (apart from quantum fluctuations), 
as can be seen from the 
comparison of the wave function patterns at different times. 
The difference between the two cases is striking.
Notice that, in this example, the localization length $\ell<1$
and one can explain the results of this simulation using 
perturbation theory. Indeed, we have $k \sim 0.35 < 1$, and 
therefore we can treat the kick $\hat U_k$ as a perturbation
of the free-evolution operatore $\hat U_T$. 
The case shown in Fig.~\ref{locfig2} is interesting since it
involves only $n=3$ qubits and a few tens on quantum gates.
Therefore this quantum computation seems to be accessible 
or close to the present capabilities of NMR-based 
\cite{NMR1,NMR2} and ion-trap \cite{ions} quantum processors.

\begin{figure}
  \includegraphics[width=9.0cm,angle=0]{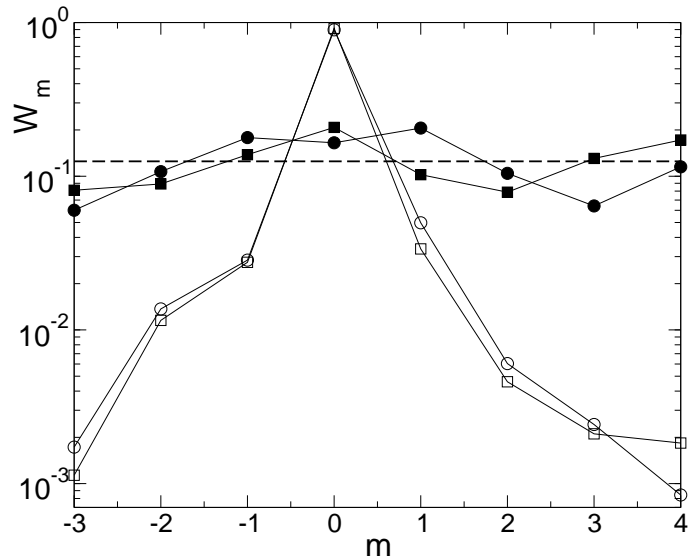}
  \caption{The probability distribution over the momentum basis for the
  sawtooth map with $n=3$ qubits, $k= K/T = K N/2 \pi L$, and initial
  momentum $m_0=0$: ergodic regime at $L=1$ (full symbols) and 
  localized regime at $L=5$ (empty symbols). 
  Circles (squares) represent the wave
  function after $t=3$ ($t=50$) time steps. 
  The dashed line represents an equally weighted wave function. 
  To smooth the results, we average over ten (one hundred) different 
  values of $K\in [1.4,1.5]$  for the localized (ergodic) case.}
  \label{locfig2}
\end{figure}

We now discuss how to extract the relevant information (the value
of the localization length) from a quantum computer simulating the
sawtooth-map dynamics. The localization length can be measured by
running the algorithm repeatedly up to time
$t>t^\star$. Each run is followed by a standard projective measurement on the
computational (momentum) basis. Since the wave function at time $t$ can be
written as
\begin{equation}
  |\psi(t)\rangle =
  \sum_m \hat\psi(m,t) \, |m\rangle \,,
\end{equation}
with $|m\rangle$ momentum eigenstates, such a measurement
gives outcome $\bar{m}$ with probability
\begin{equation}
  W_{\bar{m}} =
  \big| \langle\bar{m}|\psi(t)\rangle \big|^2 =
  \big| \hat\psi(\bar{m},t) \big|^2 \,.
\end{equation}
A first series of measurements would allow us to give a rough estimate of the
variance $\langle(\Delta{m})^2\rangle$ of the distribution
$W_m$. In turn, $\sqrt{\langle(\Delta{m})^2\rangle}$ gives a
first estimate of the localization length $\ell$. After this, we
can store the results of the measurements in histogram bins of
width $\delta{m}\propto\ell \approx
\sqrt{\langle(\Delta{m})^2\rangle}$. Finally, the localization
length is extracted from a fit of the exponential decay of this coarse-grained
distribution over the momentum basis. Elementary statistical theory tells us
that, in this way, the localization length can be obtained with accuracy $\nu$
after the order of $1/\nu^2$ computer runs. It is interesting to note
that it is sufficient to perform a coarse-grained measurement to generate a
coarse-grained distribution. This means that it will be sufficient to
measure the most significant qubits, and ignore those that would give a
measurement accuracy below the coarse graining $\delta{m}$. Thus, the number of
runs and measurements is independent of $\ell$.

In Fig.~\ref{figist}, we report a simulation of the measurement process.
In the left figure we compare the exact probabilities given by  
the wave function with the result of a complete measurement of all 
qubits and the result of a coarse-grained measurement. The histograms
are built from the same number of computational runs, each  
followed by a projective measurement. 
The coarse-grained measurement does
not resolve the thinnest structures of the exact wave function. 
However, it is still possible to extract a good estimate of
the localization length $\ell$
from a fit of the exponential decay of the probability distribution $W_m$. 
In the right figure we compare the localization lengths, extracted from the
complete and the coarse-grained measurements, as a function of the number
$N_M$ of projective measurements. Two distinct behaviors are clearly
distinguishable: the localization length computed from the 
complete measurement of all qubits converges slowly to the exact
value for the localization length, since a large number of 
projective measurement is required in order to resolve the 
exponentially decaying tails. On the contrary, the coarse-grained
measurements approaches the exact value after a much smaller number of
measurements, even though the fluctuations as a function of the number of
measurements are quite large. 

It is possible to give a better estimate 
of the localization length by computing the inverse participation ratio  
\begin{equation}
\xi = \frac{1}{\sum_m  W_m^2}.
\end{equation} 
The inverse participation ratio determines the number of basis states
significantly populated by the wave function and gives an estimate of 
the localization length of the system. 
We have $1\le \xi \le N$, with the limiting cases $\xi=1$ 
and $\xi=N$ corresponding to a wave function delta-peaked
($W_m=\delta_{m,m_0}$) or uniformly spread ($W_m=1/N$). 
In the localized regime, $\xi \approx \ell/2$. 
We stress that the inverse participation ratio 
is almost insensitive to the behavior of exponentially 
small tails of the wave function. 
Thus, the estimate $\ell\approx 2\xi$ is quite accurate
already with a small number of coarse-grained mesurement 
(see Fig.~\ref{figist}).

\begin{figure}
  \includegraphics[width=8.4cm,angle=0]{fig3a.eps}
  \hspace{0.5cm}
  \includegraphics[width=7.9cm,angle=0]{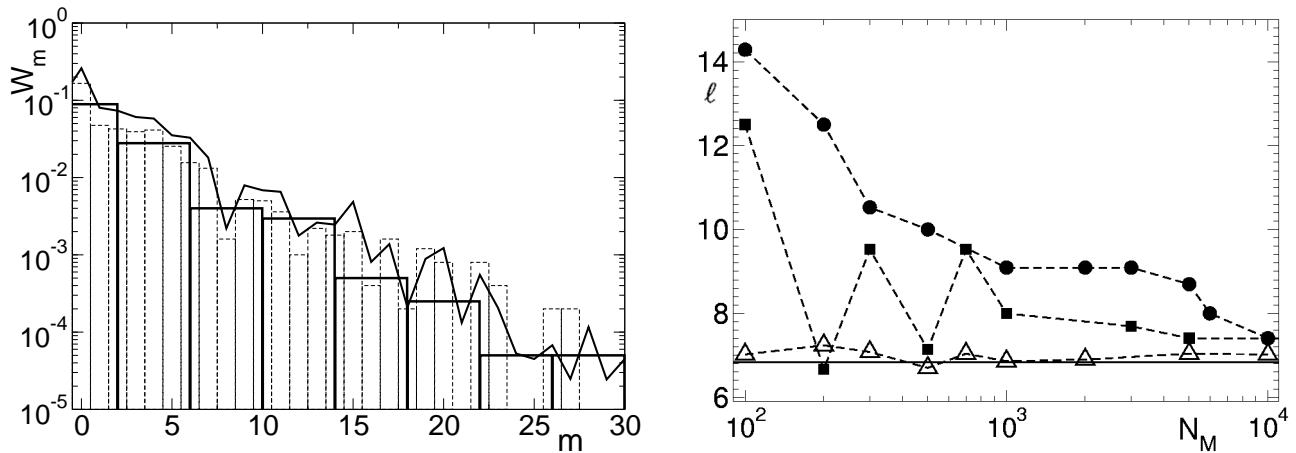}
\caption{Left: Simulation of a measurement experiment for 
  the quantum sawtooth map at $n=6$,
  $K=\sqrt{2}$, $T= 2 \pi L /N$, $L=10$, $t=50$. 
  The thick line is the exact wave function, the 
  thin dashed (thick full) histogram represents the result of 
  $N_M$ runs, each followed by a projective measurement of
  all (all except two) qubits. 
  In both cases $N_M=5 \times 10^3$. Right: The estimated 
  localization length $\ell$ as a function of the number $N_M$ of 
  projective measurements. We estimate $\ell$ by fitting the  
  probability decay for the complete (circles) and coarse-grained 
  (squares) measurements. 
  Triangles give $2\xi$, with the inverse partecipation ratio 
  computed from the coarse-grained probability distribution.
  The straight line is the theoretical result $\ell \sim 6.8$,
  obtained from Eq.~(\ref{loc2}).}
\label{figist}
\end{figure}

We now come to the crucial point, of estimating the gain of quantum
computation of the localization length with respect to classical computation.
First of all, we recall that it is necessary to make about $t^\star=O(\ell)$
map iterations to obtain the localized distribution, see
Eq.~(\ref{loc1}). This is true, both for the present
quantum algorithm and for classical computation. It is reasonable to use a
basis size $N=O(\ell)$ to detect localization (say, $N$
equal to a few times the localization length). In such a situation, 
a classical computer requires $O(\ell^2\log\ell)$
operations to extract the localization length, while a quantum computer 
would require $O(\ell(\log\ell)^2)$ elementary gates. 
Indeed, both classical and quantum computers need to  
perform $t\approx t^\star=O(\ell)=O(N)$ map iterations.
Therefore, the quantum computer provides a 
\emph{quadratic speed up} in computing the localization length, 
As we shall see in Sec.~\ref{sec:fidelity}, 
the quantum computation can provide an \emph{exponential gain} 
(with respect to any known classical computation) in problems 
that require the simulation of dynamics up to a time $t$ which is 
independent of the number of qubits. In this case, provided that 
we can extract the relevant information in a number of measurements 
polynomial in the number of qubits, one should compare 
$O(t(\log{N})^2)$ elementary gates (quantum computation)
with $O(tN\log{N})$ elementary gates (classical computation).

\section{Quantum computing of Brownian and anomalous diffusion}
\label{sec:anomalous}

As we have discussed in Sec.~\ref{sec:sawmap}, the classical sawtooth map 
is characterized by different diffusive behaviors in the chaotic 
and semi-integrable regimes. Quantum computers could  
help us to study these different regimes by simulating the map 
in the deep semiclassical region $\hbar_\text{eff}\to 0$. 
Let us first show that a quantum computer would be useful in computing 
the Brownian diffusion coefficient $D_m$.
For this purpose, we can repeat several times the quantum simulation 
of the sawtooth map up to a given time $t$, ending each 
run with a standard projective measurement in the momentum 
basis. This allows us to compute, up to statistical 
errors, $\langle(\Delta{m})^2\rangle$.
The diffusion coefficient is then obtained from Eq.~(\ref{siberia2})
as $D_m\approx \langle(\Delta{m})^2\rangle/t$. 
Therefore a computation of the diffusion coefficient up 
to time $t$ significantly involves the order of 
$\sqrt{D_m t}$ momentum eigenstates 
(other levels are only weakly populated for times 
smaller than $t$ and can be neglected).
Thus, a basis of dimension $N=O(t^{1/2})$
is sufficient for this computation. To estimate the
speed up of quantum computation, one should compare 
$O(t(\log{N})^2)=O(N^2(\log{N})^2)$ elementary gates 
(quantum computation) with 
with $O(tN\log{N})=O(N^3\log(N))$ elementary gates 
(classical computation). This gives an {\it algebraic 
speed up}.

We note that similar computations could be 
done in the regime of anomalous diffusion, in which
$\langle (\Delta J)^2\rangle = 
T^2 \langle (\Delta m)^2\rangle \propto t^\alpha$, to 
evaluate the exponent $\alpha$, a quantity of 
great physical interest. Such a regime is quite
complex in the sawtooth map: 
Fig.~\ref{fanom} shows, for the classical map, 
the dependence of the exponent $\alpha$
as a function of $K$. As can be seen from this figure,
the map explores subdiffusive ($\alpha<1$) and superdiffusive 
($\alpha>1$) regions, up to ballistic diffusion ($\alpha=2$).
As required by the principle of quantum to classical 
correspondence, the quantum  sawtooth map follows this behavior 
in the deep semiclassical regime $\hbar_\text{eff}\ll 1$,
up to some time scale which diverges when $\hbar_\text{eff}\to 0$.
It is important to point out that $\hbar_\text{eff}$
drops to zero exponentially with the number of qubits
($\hbar_\text{eff}\propto 1/N=1/2^{n}$), and therefore 
the deep semiclassical region can be reached with a small number
of qubits.
For large $\hbar_\text{eff}$, one can also study how diffusion
is modified by important quantum phenomena, like quantum tunneling,
localization, and quantum resonances.

A quantum computer could help us in obtaining the 
exponent $\alpha$ of the anomalous diffusion.
In this case, since $\langle (\Delta m)^2 \rangle \propto t^\alpha$, 
a rough estimate 
of the size of the basis required for the computation up 
to time $t$ is $N=O(t^{\alpha/2})$. Hence, we 
must compare 
$O(t(\log{N})^2)=O(N^{2/\alpha}(\log{N})^2)$ elementary gates 
(quantum computation) with 
$O(tN\log{N})=O(N^{(\alpha+2)/\alpha}\log(N))$ elementary gates 
(classical computation). The speed up is again algebraic.

\begin{figure}
\includegraphics[width=9.0cm,angle=0]{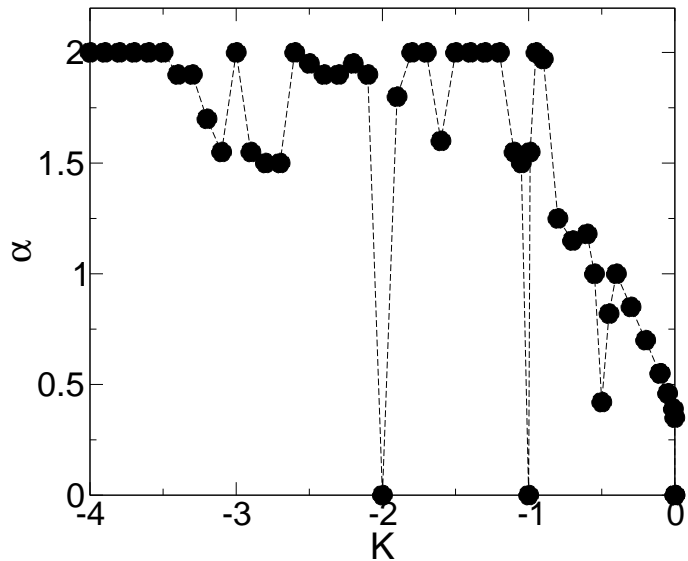}
\caption{Left: Exponent $\alpha$ of the anomalous diffusion 
($\langle (\Delta J)^2\rangle \propto t^\alpha$) as a function of $K$ 
for the classical sawtooth map in the semi-integrable regime.} 
\label{fanom}
\end{figure}

\section{Quantum computing of the fidelity of quantum motion}
\label{sec:fidelity}

The simulation of quantum dynamics up to a time $t$
which is independent of the number of qubits is useful, for
instance, to measure dynamical correlation functions of the form
\begin{equation}
  C(t) \equiv
  \langle\psi| \, \hat{A}^\dagger(t) \, \hat{B}(0) \, |\psi\rangle =
  \langle\psi| \, (\hat{U}^\dagger)^t \, \hat{A}^\dagger(0) 
  \, \hat{U}^t \, \hat{B}(0) \, |\psi\rangle
  \,,
  \label{corrfun}
\end{equation}
where $\hat{U}$ is the time-evolution operator (\ref{sawq}) for 
the sawtooth map. Similarly, we can efficiently compute
the fidelity of quantum motion, 
which is a quantity of central interest in the study of the stability
of a system under perturbations (see, e.g., 
\cite{peres,jalabert,jacquod,tomsovic,PRE,prosen,emerson,zurek} and 
references therein). 
The fidelity $f(t)$ (also called
the Loschmidt echo), measures the accuracy with which a quantum
state can be recovered by inverting, at time $t$, the dynamics with a perturbed
Hamiltonian. It is defined as
\begin{equation}
  f(t) = \langle\psi| \, (\hat{U}^\dagger_\epsilon)^t \, 
\hat{U}^t \, |\psi\rangle \,.
\end{equation}
Here the wave vector $|\psi\rangle$ evolves forward in time with the
Hamiltonian $H$ of Eq.~(\ref{sawq}) up to time $t$, and then evolves backward
in time with a perturbed Hamiltonian $H_\epsilon$ 
($\hat{U}_\epsilon$ is the
corresponding time-evolution operator). For instance, we can perturb
the parameter $k$ in the sawtooth map as follows: $k\to{}k^\prime=k+\epsilon$,
with $\epsilon\ll k$.
If the evolution operators $\hat{U}$ and 
$\hat{U}_\epsilon$ can be simulated efficiently on
a quantum computer, as is the case in most physically interesting
situations, then the fidelity of quantum motion can be evaluated with
exponential speed up with respect to known classical
computations. The same conclusion is valid for the correlation functions
(\ref{corrfun}).

The fidelity can be efficiently evaluated on a quantum computer, 
with the only requirement of an ancilla qubit, using the 
\emph{scattering circuit} drawn in Fig.~\ref{scattering}
\cite{zoller,saraceno}. 
This circuit has various important applications in quantum computing, 
including quantum state tomography and quantum spectroscopy
\cite{saraceno}. The circuit ends up with the measurement of 
the ancilla qubit, and we have 
\begin{equation}
\langle \sigma_z \rangle = \mathrm{Re} [\mathrm{Tr} (\hat{W} \rho)], \quad
\langle \sigma_y \rangle = \mathrm{Im} [\mathrm{Tr} (\hat{W} \rho)],
\label{scatt} 
\end{equation}
where $\langle \sigma_z \rangle$ and $\langle \sigma_y \rangle$ are
the expectation values of the Pauli spin operators $\hat{\sigma}_z$ and
$\hat{\sigma}_y$ for the ancilla qubit, and $\hat{W}$ is a unitary operator.
These two expectation values can be obtained (up to statistical errors)
if one runs several times the scattering circuit.
If we set $\rho= |\psi \rangle \langle \psi |$ and
$\hat{W} = (\hat{U}_{\epsilon}^{\dagger})^t \, \hat{U}^{t}$, it is 
easy to see that $f(t) = | \mathrm{Tr} ( \hat{W} \rho) |^{2}=
\langle \sigma_z \rangle^2 + \langle \sigma_y \rangle^2$.
For this reason, provided the quantum algorithm which implements $\hat{U}$ is
efficient, as it is the case for the quantum sawtooth map, the fidelity can 
be efficiently computed by means of the circuit described
in Fig.~\ref{scattering}.
We note that another possible way to efficiently measure the 
fidelity has been proposed in \cite{emerson}.

\begin{figure}
\includegraphics[width=13.0cm,angle=0]{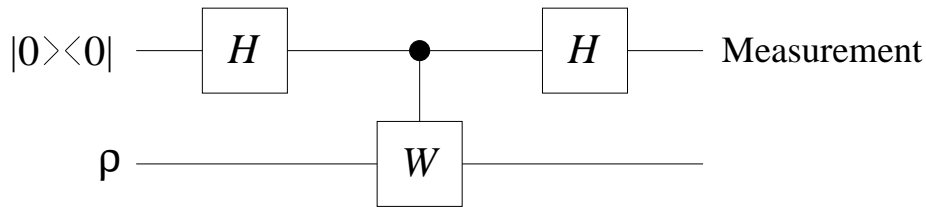}
\caption{Scattering circuit. The top line denotes the 
ancilla qubit, the bottom line a set of $n$ qubits, $H$ 
the Hadamard gate, and $W$ a unitary transformation.}
\label{scattering}
\end{figure}

\section{Conclusions}
\label{sec:conclusions}

In this paper, we have discussed relevant physical 
examples of efficient 
information extraction in the quantum computation of 
a dynamical system.
We have shown that a quantum computer with a small number
of qubits can efficiently simulate the quantum localization
effects, simulate both the Brownian and anomalous diffusion in the 
deep semiclassical regime, and compute the fidelity of quantum motion. 
We would like to stress that the simulation of complex dynamical 
systems is accessible to the first generation of quantum 
computers with less than 10 qubits. Therefore, we believe
that quantum algorithms for dynamical systems deserve further 
studies, since they are the ideal software for the first
quantum processors. Furthermore, we emphasize that the quantum 
computation of quantities like dynamical localization 
or fidelity is a demanding testing ground for quantum 
computers. In the first case, we want to simulate dynamical 
localization, a purely quantum phenomena which is quite 
fragile in the presence of noise \cite{ott,song}; in the latter case,
fidelity is computed as a result of a sophisticated 
many-qubit Ramsey-type interference experiment. 
Therefore the computation of these quantities appears
to be a relevant test for quantum processors 
operating in the presence of decoherence and imperfection 
effects.

\begin{acknowledgments}

This work was supported in part by the EC contracts 
IST-FET EDIQIP and RTN QTRANS, the NSA and ARDA under
ARO contract No. DAAD19-02-1-0086, and the PRIN 2002 
``Fault tolerance, control and stability in
quantum information processing''.

\end{acknowledgments}

\bibliographystyle{prsty}

\begin{thebibliography}{99}
\bibitem{footnote} Strictly speaking, the asymptotic tails of the 
localized wave functions decay, for the sawtooth map model, 
as a power law:
\begin{equation}
W_m\propto \frac{1}{|m-m_0|^4}.
\label{power}
\end{equation}
This happens due to the discontinuity in the kicking force 
$F(\theta)=k(\theta-\pi)$, when the angle variable
$\theta=0$. For this reason the matrix elements of
the evolution operator $\hat{U}$ [defined by Eq.~(\ref{sawq})] 
decay as a power law in the momentum eigenbasis:
$U_{mm^\prime}=\langle m|\hat{U}|m^\prime \rangle
\sim 1/|m-m^\prime|^\alpha$, with $\alpha=2$.
This case was investigated for random matrices,
where it was shown that eigenfunctions are also
algebraically localized with the same exponent
$\alpha$ \cite{brm}.
However, the localization picture is not very
sensitive to the behavior of the tails of the wave
function. Indeed, a rough estimate of the crossover
between the exponential decay (\ref{expdecay}) and
the power law decay (\ref{power}) is given by their
crossing point,
\begin{equation}
m_c \sim \frac{3}{2} \ell \log \ell, \quad
W_m(m_c)\sim \frac{1}{\ell^4\log\ell}.
\end{equation}
This implies that by increasing $\ell$ the exponential
localization is pushed to larger momentum windows and
down to smaller probabilities.
\end{thebibliography}

\end{document}